\documentclass[conference,compsoc]{IEEEtran}

\usepackage[nocompress]{cite}
\usepackage{graphicx}
\usepackage{float}
\usepackage{tabularx}
\usepackage[hidelinks]{hyperref}
\usepackage{booktabs}
\usepackage{enumitem}
\usepackage{rotating}
\usepackage{amssymb}
\usepackage{svg}

\begin{document}
\title{SoK: Web3 RegTech for Cryptocurrency VASP AML/CFT Compliance}

\author{
\IEEEauthorblockN{Qian'ang Mao, Jiaxin Wang, Ya Liu, Li Zhu, Jiaman Chen, Jiaqi Yan}
\IEEEauthorblockA{Nanjing University, Nanjing, China}
}

\newcommand{\fullcirc}{$\bullet$}
\newcommand{\halfcirc}{$\circ$}
\newcommand{\emptycirc}{---}

\maketitle

\begin{abstract}
The decentralized architecture of Web3 technologies creates fundamental challenges for Anti-Money Laundering and Counter-Financing of Terrorism compliance. Traditional regulatory technology solutions designed for centralized financial systems prove inadequate for blockchain's transparent yet pseudonymous networks. This systematization examines how blockchain-native RegTech solutions leverage distributed ledger properties to enable novel compliance capabilities.

We develop three taxonomies organizing the Web3 RegTech domain: a regulatory paradigm evolution framework across ten dimensions, a compliance protocol taxonomy encompassing five verification layers, and a RegTech lifecycle framework spanning preventive, real-time, and investigative phases. Through analysis of 41 operational commercial platforms and 28 academic prototypes selected from systematic literature review (2015-2025), we demonstrate that Web3 RegTech enables transaction graph analysis, real-time risk assessment, cross-chain analytics, and privacy-preserving verification approaches that are difficult to achieve or less commonly deployed in traditional centralized systems.

Our analysis reveals critical gaps between academic innovation and industry deployment, alongside persistent challenges in cross-chain tracking, DeFi interaction analysis, privacy protocol monitoring, and scalability. We synthesize architectural best practices and identify research directions addressing these gaps while respecting Web3's core principles of decentralization, transparency, and user sovereignty.
\end{abstract}

\begin{IEEEkeywords}
Cryptocurrency, Compliance, RegTech, Anti-Money Laundering, Counter-Financing of Terrorism, Know Your Customer, Know Your Transaction, Address Screening
\end{IEEEkeywords}

\IEEEpeerreviewmaketitle

\section{Introduction}

The rapid proliferation of the Web3 ecosystem and cryptocurrencies has ushered in a new era of digital markets, bringing unprecedented opportunities for innovation, efficiency, and financial inclusion \cite{bonneau2015sokresearchperspectives,voshmgir2020tokeneconomyhow,abdelhamid2025reviewblockchaintechnology}. Yet the very attributes that empower these technologies, including decentralization, pseudonymity, and global accessibility, simultaneously pose profound challenges to Anti-Money Laundering (AML) and Countering the Financing of Terrorism (CFT) compliance \cite{force2021updatedguidanceriskbased,financialactiontaskforcefatf2012internationalstandardscombating,huang2020antimoneylaunderingblockchain}. The inherent characteristics of blockchain systems have exposed the limitations of traditional compliance frameworks, which were primarily designed for centralized financial infrastructures, rendering them ill-suited for the complexities of the digital asset landscape \cite{weber2019antimoneylaunderingbitcoin,pocher2023detectinganomalouscryptocurrency,bankforinternationalsettlementsbis2022futuremonetarysystem,aldasoroapproachantimoneylaundering}. In response, a new generation of regulatory technology (RegTech) solutions has emerged, purpose-built to address the distinct compliance demands of the Web3 environment \cite{arner2016fintechregtechreconceptualization,charoenwong2024regtechtechnologydrivencompliance,prosper2025blockchainbasedsolutionsantimoney}.

Despite significant progress in blockchain compliance capabilities, formidable challenges remain. The cryptocurrency compliance landscape exists in a constant state of flux, with illicit actors and compliance providers locked in perpetual arms race dynamics \cite{albrecht2019usecryptocurrenciesmoney,guidara2022cryptocurrencymoneylaundering,prosper2025blockchainbasedsolutionsantimoney}. Novel evasion tactics emerge rapidly as adversaries adapt to new detection methods, requiring RegTech solutions to maintain continuous innovation cycles \cite{unitednationsofficeondrugsandcrimemoneylaunderingcryptocurrencies}. Compliance providers respond by developing advanced analytical capabilities leveraging graph neural networks, AI-powered behavioral analytics, and sophisticated forensic techniques \cite{azad2025machinelearningblockchain,rodriguezvalencia2025systematicreviewartificial,zebari2025comprehensivereviewintegrating}. This adversarial co-evolution shows no signs of abating, necessitating adaptive RegTech solutions that can evolve alongside emerging threats \cite{verma2024impactcryptocurrencymoney,tiwari2025understandingevolvinglandscape}.

The absence of universally accepted standards poses significant challenges across multiple compliance domains. Risk scoring methodologies for addresses, transactions, and entities lack standardization, hindering Virtual Asset Service Providers' ability to benchmark compliance efforts and limiting regulators' capacity to compare different approaches objectively. The FATF Travel Rule mandates information exchange between VASPs, yet the absence of a single universally adopted protocol for secure and interoperable data sharing has resulted in fragmented solutions and operational complexities \cite{force2021updatedguidanceriskbased,pocher2022privacytransparencycbdcs}. Rapid blockchain innovation continually introduces new protocols, token standards, Layer-2 solutions, and decentralized applications, creating moving targets for compliance tools \cite{mohammedabdul2024navigatingblockchainstwin,xu2024exploringblockchaintechnology,bankforinternationalsettlementsbis2022futuremonetarysystem}. Each technological advancement introduces unique risks requiring constant analytical adaptation. The growing adoption of DeFi protocols and self-custody wallets further complicates compliance by limiting VASP visibility into customer activities, creating blind spots for AML/CFT monitoring and Travel Rule implementation when funds move off-platform \cite{harvey2021defifuturefinance,salami2021challengesapproachesregulating}.

\subsection{Research Questions and Contributions}

This paper systematizes knowledge about Web3 RegTech solutions and the regulatory compliance lifecycle. Information about these systems has been scattered across industry whitepapers, academic publications, regulatory guidance documents, and practitioner reports. We provide foundational background on compliance regulation, analyze current solutions across industry and academia, and identify critical gaps and research directions. We organize our contributions around three research questions.

\noindent\textbf{RQ1: What are the fundamental architectural differences between Web3 and traditional RegTech solutions, and what novel technical capabilities do blockchain-native approaches enable for VASP AML/CFT compliance?}
We systematically compare centralized and blockchain-native compliance architectures across seven dimensions: data accessibility, verification mechanisms, identity models, monitoring scope, analytical approaches, enforcement models, and transparency guarantees \cite{arner2016fintechregtechreconceptualization,zetsche2017fintechtechfinregulatory}. Our analysis demonstrates that Web3 RegTech constitutes a paradigmatic shift rather than incremental evolution, enabling comprehensive transaction graph analysis \cite{meiklejohn2013fistfulbitcoinscharacterizing,reid2013analysisanonymitybitcoin}, real-time risk assessment \cite{moser2014riskscoringbitcoin,weber2019antimoneylaunderingbitcoin}, cross-chain analytics \cite{yousaf2019tracingtransactionscryptocurrency,zamyatin2021sokcommunicationdistributed}, smart contract interaction analysis \cite{chen2020understandingethereumgraph,qin2021attackingdefiecosystem}, and privacy-preserving verification \cite{ben-sasson2017scalablezeroknowledge,zhang2024blockchainzeroknowledge}. We contextualize these architectural differences with recent advances in consensus diagnostics, hybrid security, and AI-optimized blockchain controls that increasingly intersect with compliance requirements \cite{ahn2024blockchainconsensusmechanisms,cai2025blockchainconsensusalgorithm,wei2024enhancedconsensusalgorithm,venkatesan2024blockchainsecurityenhancement,yuan2025aidrivenoptimizationblockchain}. Through examination of 41 operational RegTech platforms (34 Web3-native and 7 traditional with blockchain capabilities), we document rapid market growth with numerous Web3-native solutions emerging since 2023. \autoref{sec:systematization} and \autoref{sec:ecosystem} address this research question.

\noindent\textbf{RQ2: What is the current landscape of Web3 RegTech across industry and academia, what are their technical approaches, and what gaps exist between proposed solutions and practical VASP compliance needs?}
We analyze 41 commercial Web3 RegTech platforms and 28 academic research prototypes, categorizing them into five functional domains: transaction monitoring \cite{weber2019antimoneylaunderingbitcoin,petterssonruiz2022combatingmoneylaundering}, risk scoring \cite{moser2014riskscoringbitcoin,monamo2016unsupervisedlearningrobust}, blockchain forensics \cite{meiklejohn2013fistfulbitcoinscharacterizing,yousaf2019tracingtransactionscryptocurrency}, compliance automation \cite{charoenwong2024regtechtechnologydrivencompliance}, and privacy-preserving verification \cite{ben-sasson2017scalablezeroknowledge,yang2024researchidentitydata}. We examine technical building blocks including graph neural networks \cite{wu2023tracerscalablegraphbased,azad2025machinelearningblockchain,asiri2025graphconvolutionnetwork,zhang2024graphnetworkmodels}, clustering algorithms \cite{reid2013analysisanonymitybitcoin,ron2013quantitativeanalysisfull}, machine learning models \cite{pourhabibi2020frauddetectionsystematic,rodriguezvalencia2025systematicreviewartificial}, heuristic detection \cite{difrancescomaesa2018datadrivenanalysisbitcoin}, and cryptographic protocols \cite{zhang2024blockchainzeroknowledge,camenisch2001efficientsystemnontransferable}. Our gap analysis identifies eight critical disconnects between academic innovation and industry adoption: technology readiness limitations, regulatory alignment challenges, data availability constraints, adversarial robustness gaps, cost-benefit misalignments, standardization deficiencies, operational workflow integration failures, and validation benchmarking inadequacies. We document six persistent challenges limiting current capabilities across cross-chain tracking \cite{zamyatin2021sokcommunicationdistributed,lin2024crosschainabnormaltransaction}, DeFi interaction analysis \cite{qin2021attackingdefiecosystem,sun2025ethereumfrauddetection}, privacy protocol analysis \cite{kappos2018empiricalanalysisanonymity,kumar2017traceabilityanalysismoneros}, real-time scalability, attribution database maintenance, and false positive management \cite{alotibi2022moneylaunderingdetection}. \autoref{sec:ecosystem} addresses this research question.

\noindent\textbf{RQ3: Based on identified architectural differences and ecosystem gaps, what are the critical research directions and best practices to advance Web3 RegTech effectiveness?}
Building on comprehensive analysis of current capabilities and limitations, we identify critical research directions encompassing verifiable compliance proofs, semantic transaction understanding, behavior-based risk assessment, privacy-preserving collaborative analytics, intent-aware monitoring systems, cross-chain attribution techniques, adversarial robustness enhancements, standardized evaluation frameworks, and regulatory-compatible privacy preservation. We present architectural best practices for scalable monitoring, privacy-computation trade-offs, and intent interpretation frameworks that developers, protocol designers, and researchers can leverage to build solutions enhancing compliance effectiveness without compromising Web3's core principles of decentralization, transparency, and user sovereignty. \autoref{sec:discussion} addresses this research question.

\subsection{Research Scope and Methodology}
This systematization draws on analysis of 41 operational RegTech platforms (34 Web3-native, 7 traditional with blockchain capabilities) and 28 academic research prototypes from systematic literature review (2015-2025). Following established methodology for technology ecosystem analysis \cite{arner2016fintechregtechreconceptualization,bonneau2015sokresearchperspectives}, we synthesize commercial platform capabilities through technical documentation, whitepapers, and public capability descriptions, complemented by academic literature providing independent validation. This mixed-methods approach enables comprehensive domain coverage while acknowledging the inherent challenges of analyzing rapidly evolving technology ecosystems where platform capabilities continuously advance.



\section{Background}

This section provides essential technical background on blockchain technology and Anti-Money Laundering/Countering the Financing of Terrorism regulatory frameworks necessary for understanding Web3 RegTech systems.

\subsection{Blockchain and Web3 Fundamentals}

Blockchain technology fundamentally restructures data integrity and transaction validation through distributed consensus mechanisms. First introduced by Nakamoto \cite{nakamoto2008bitcoinpeertopeerelectronic}, blockchain maintains a continuously growing ledger of cryptographically linked blocks, each containing timestamped transaction data and a reference to its predecessor. This architecture creates an immutable record that can be independently verified by any network participant, eliminating reliance on centralized authorities for transaction validation.

Web3 represents the evolution of internet infrastructure toward decentralized architectures built atop blockchain foundations \cite{wood2014ethereumsecuredecentralised}. Smart contracts constitute the primary programmability layer, enabling self-executing code that automatically enforces agreement terms without intermediary oversight \cite{szabo1997formalizingsecuringrelationships}. Decentralized Finance protocols leverage smart contracts to recreate traditional financial services including lending markets, trading venues, and derivative instruments entirely through algorithmic intermediation \cite{harvey2021defifuturefinance}. Decentralized Autonomous Organizations encode governance structures in smart contracts, enabling collective decision-making through token-weighted voting mechanisms \cite{voshmgir2020tokeneconomyhow}. Non-Fungible Tokens provide cryptographically secured ownership records for unique digital or physical assets \cite{wang2021nonfungibletokennft}.

The transparency inherent in public blockchains fundamentally differs from traditional financial systems. Every transaction broadcasts to all network participants, creating comprehensive visibility into fund flows and contract interactions \cite{narayanan2016bitcoincryptocurrencytechnologies}. This transparency enables novel analytical approaches while simultaneously creating privacy challenges, as the permanent public record of all activities can reveal sensitive information when addresses become associated with real-world identities.

\subsection{AML/CFT Regulatory Framework}

Anti-Money Laundering and Countering the Financing of Terrorism regulations establish baseline obligations that financial institutions must satisfy to prevent illicit use of financial systems \cite{financialactiontaskforcefatf2012internationalstandardscombating}. Customer Due Diligence requires verification of customer identities and understanding the nature of business relationships, enabling institutions to assess appropriate risk levels and design proportionate monitoring strategies. Transaction Monitoring demands continuous surveillance of customer activities to identify patterns inconsistent with known legitimate business or personal activities. Suspicious Activity Reporting mandates that institutions notify authorities when transactions potentially indicate money laundering, terrorist financing, fraud, or other criminal activities. Record Keeping obligations require maintenance of identification data and transaction records for specified periods, typically five years, ensuring availability for retrospective investigation.

The Financial Action Task Force, an intergovernmental organization, establishes global standards for AML/CFT compliance \cite{financialactiontaskforcefatf2012internationalstandardscombating}. FATF's Recommendation 16, commonly known as the Travel Rule, requires Virtual Asset Service Providers to obtain, hold, and exchange information about originators and beneficiaries of virtual asset transfers \cite{force2021updatedguidanceriskbased}. Specifically, VASPs must transmit originator's name, account number or address, and beneficiary's name and account number for transfers exceeding specified thresholds. This requirement attempts to replicate wire transfer transparency in cryptocurrency contexts, though technical implementation proves significantly more challenging due to decentralized network architectures and interoperability gaps among Travel Rule protocols \cite{force2021opportunitieschallengesnew}.

Virtual Asset Service Providers, as defined by FATF, are entities conducting activities or operations on behalf of customers including exchange between virtual and fiat currencies, transfer of virtual assets, safekeeping or administration of virtual assets, and participation in financial services related to virtual assets \cite{force2021updatedguidanceriskbased}. VASPs face compliance obligations similar to traditional financial institutions but must address unique challenges posed by blockchain technology including pseudonymous addresses lacking inherent identity information, cross-chain transactions challenging single-ledger monitoring approaches, self-custody eliminating VASP visibility for off-platform activities, and smart contract complexity creating interpretation challenges.

\subsection{Compliance Challenges in Web3}

The architectural characteristics of Web3 systems create fundamental tensions between regulatory compliance requirements and technological design principles. Privacy-enhancing technologies provide legitimate financial privacy protections but simultaneously enable obfuscation of illicit fund flows. Coin mixing services, privacy-focused cryptocurrencies like Monero and Zcash, and zero-knowledge proof systems all complicate transaction tracing and risk assessment \cite{moser2018empiricalanalysistraceability,kappos2018empiricalanalysisanonymity}. Recent research explores privacy-preserving compliance approaches balancing these competing interests through selective disclosure mechanisms, decentralized identifiers, and zero-knowledge proofs of compliance \cite{yang2024researchidentitydata,ben-sasson2017scalablezeroknowledge,zhang2024blockchainzeroknowledge}.

Decentralization challenges traditional accountability frameworks that assume identifiable intermediaries responsible for compliance. Decentralized Finance protocols execute autonomously through smart contracts without ongoing human intervention, raising questions about who bears compliance responsibility \cite{harvey2021defifuturefinance,werbach2018trustverifywhy}. Protocol developers may have limited control after deployment, liquidity providers supply capital without operational control, and users interact directly with contracts without intermediary oversight. This disintermediation creates regulatory gaps where no clearly accountable party exists to enforce compliance requirements.

Technological innovation proceeds at a pace that regulatory frameworks and compliance tools struggle to match. New token standards, novel DeFi mechanisms, Layer-2 scaling solutions, and cross-chain bridges continuously introduce unprecedented interaction patterns requiring compliance interpretation \cite{mohammedabdul2024navigatingblockchainstwin,xu2024exploringblockchaintechnology}. Each innovation potentially introduces new risks or evasion vectors that existing detection systems fail to recognize, necessitating continual model retraining and intelligence sharing.

\section{Systematization of Web3 RegTech}
\label{sec:systematization}

\begin{figure*}[t]
\centering
\includegraphics[width=0.95\linewidth]{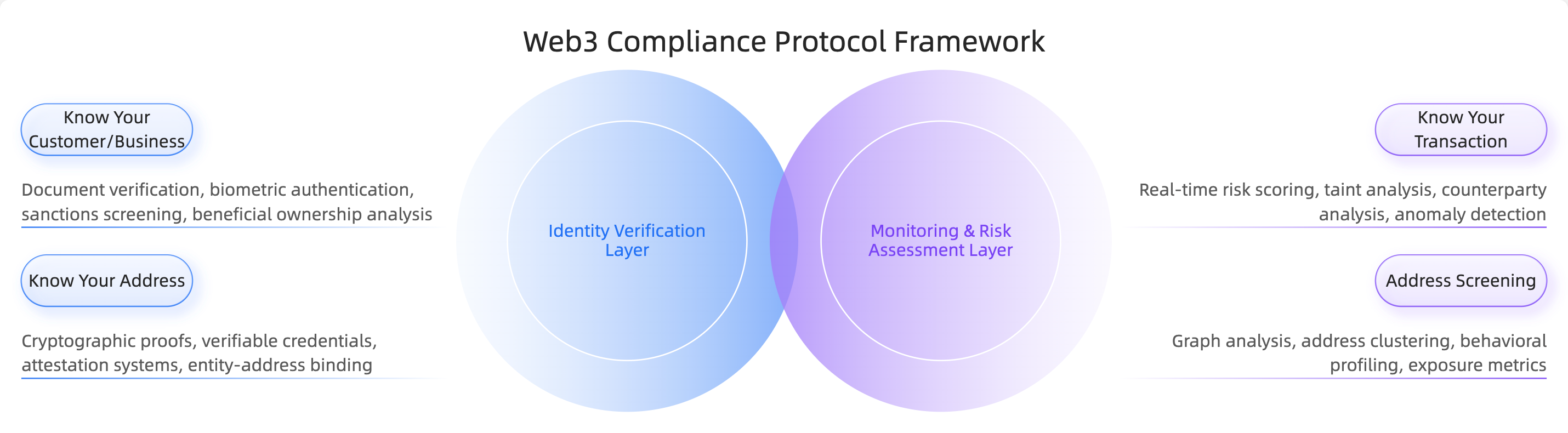}
\caption{Web3 Compliance Protocol Framework: Two-Layer Architecture. This framework organizes compliance protocols into two fundamental layers: (1) Identity Verification Layer establishes and verifies identities across entity level (Know Your Customer/Business) and address level (Know Your Address) through document verification, biometric authentication, sanctions screening, and cryptographic proofs; (2) Monitoring \& Risk Assessment Layer provides continuous surveillance and risk evaluation through transaction monitoring (Know Your Transaction) and composite address risk scoring (Address Screening) using real-time risk scoring, taint analysis, graph analysis, and behavioral profiling.}
\label{fig:framework}
\end{figure*}

This section presents the core systematization of Web3 RegTech knowledge, as shown in \autoref{fig:framework}, establishing taxonomies and frameworks that organize this rapidly evolving domain. 

\subsection{Regulatory Paradigm Evolution}

The regulatory landscape has evolved through three distinct paradigms, each characterized by fundamentally different data architectures, verification mechanisms, and enforcement approaches. Table \ref{tab:regulatory_evolution} systematizes these differences across ten critical dimensions, demonstrating that Web3 RegTech constitutes a necessary adaptation rather than optional enhancement.

\begin{table*}[t]
\centering
\caption{Evolution of Regulatory Paradigms: Multi-Dimensional Comparison}
\label{tab:regulatory_evolution}
\scriptsize
\begin{tabularx}{\textwidth}{lXXX}
\toprule
\textbf{Dimension} & \textbf{Traditional Regulation} & \textbf{Virtual Asset Regulation (Centralized)} & \textbf{Web3 Regulation (Decentralized)} \\
\midrule
\textbf{Data Architecture} & Centralized institutional databases with bilateral access controls \cite{arner2016fintechregtechreconceptualization} & Centralized VASP databases with enhanced cross-border reporting \cite{force2021updatedguidanceriskbased} & Distributed public ledgers with universal read access \cite{nakamoto2008bitcoinpeertopeerelectronic,buterin2014ethereumnextgenerationsmart} \\
\midrule
\textbf{Transaction Visibility} & Limited to single institution's customer relationships \cite{zetsche2017fintechtechfinregulatory} & Extended to VASP's complete customer base with Travel Rule exchange \cite{force2021opportunitieschallengesnew} & Network-wide visibility across all participants and intermediaries \cite{meiklejohn2013fistfulbitcoinscharacterizing,reid2013analysisanonymitybitcoin} \\
\midrule
\textbf{Identity Foundation} & Government-issued documents verified through physical or digital submission \cite{financialactiontaskforcefatf2012internationalstandardscombating} & Same document-based approach with enhanced digital verification \cite{charoenwong2024regtechtechnologydrivencompliance} & Cryptographic identifiers with optional verifiable credential attestation \cite{reed2022decentralizedidentifiersdids,camenisch2001efficientsystemnontransferable} \\
\midrule
\textbf{Verification Model} & Trust in institutional controls and periodic audits \cite{arner2016fintechregtechreconceptualization} & Trust in VASP compliance plus regulatory oversight \cite{force2021updatedguidanceriskbased} & Cryptographic proof verification without trust assumptions \cite{ben-sasson2017scalablezeroknowledge,zhang2024blockchainzeroknowledge} \\
\midrule
\textbf{Intermediary Role} & Mandatory for transaction execution and record-keeping \cite{financialactiontaskforcefatf2012internationalstandardscombating} & Required for fiat-crypto conversion, optional for crypto-crypto \cite{trautman2014virtualcurrenciesbitcoin} & Optional or absent, particularly in DeFi and self-custody scenarios \cite{harvey2021defifuturefinance,qin2021attackingdefiecosystem} \\
\midrule
\textbf{Monitoring Approach} & Rule-based transaction surveillance within institutional silos \cite{drezewski2015applicationsocialnetwork} & Rule-based plus behavioral analytics within VASP ecosystem \cite{weber2019antimoneylaunderingbitcoin} & Graph-based behavioral analytics across entire transaction network \cite{wu2023tracerscalablegraphbased,azad2025machinelearningblockchain} \\
\midrule
\textbf{Risk Assessment Basis} & Customer profile, transaction patterns, jurisdiction \cite{financialactiontaskforcefatf2012internationalstandardscombating} & Enhanced with on-chain activity history and blockchain forensics \cite{moser2014riskscoringbitcoin} & Multi-dimensional including address clustering, taint analysis, protocol risk \cite{moser2013inquirymoneylaundering,difrancescomaesa2018datadrivenanalysisbitcoin} \\
\midrule
\textbf{Enforcement Mechanism} & Account freezes, transaction blocks, license revocation \cite{financialactiontaskforcefatf2012internationalstandardscombating} & Similar to traditional plus blockchain address blacklisting \cite{force2021updatedguidanceriskbased} & Smart contract restrictions, protocol-level compliance, limited retroactive control \cite{szabo1997formalizingsecuringrelationships,luu2016makingsmartcontracts} \\
\midrule
\textbf{Cross-Border Complexity} & High, requiring multi-jurisdictional coordination \cite{financialactiontaskforcefatf2012internationalstandardscombating} & Very high, compounded by pseudonymity and rapid value transfer \cite{force2021opportunitieschallengesnew,salami2021challengesapproachesregulating} & Extreme, with permissionless global access and regulatory arbitrage \cite{bonneau2015sokresearchperspectives,werbach2018trustverifywhy} \\
\midrule
\textbf{Privacy-Compliance Tension} & Moderate, managed through data protection regulations \cite{zetsche2017fintechtechfinregulatory} & Significant, balancing transparency needs with privacy rights \cite{pocher2022privacytransparencycbdcs} & Fundamental, as public ledgers expose transaction histories permanently \cite{androulaki2013evaluatinguserprivacy,koshy2014analysisanonymitybitcoin} \\
\bottomrule
\end{tabularx}
\end{table*}

Traditional financial regulation emerged around centralized institutional architectures where financial intermediaries maintained exclusive control over customer data and transaction records \cite{arner2016fintechregtechreconceptualization,zetsche2017fintechtechfinregulatory}. Regulators accessed this information through reporting requirements and examination authority, creating hierarchical oversight models with clear accountability chains. Customer Due Diligence relied on document submission and human review, with verification quality depending on institutional processes. Transaction monitoring employed rule-based systems detecting threshold violations within single-institution data silos \cite{drezewski2015applicationsocialnetwork}.

Virtual Asset Service Provider regulation attempted to extend traditional frameworks to cryptocurrency intermediaries. The Financial Action Task Force expanded its recommendations to encompass VASPs, requiring Know Your Customer verification, transaction monitoring, and suspicious activity reporting comparable to traditional financial institutions \cite{financialactiontaskforcefatf2012internationalstandardscombating,force2021updatedguidanceriskbased}. The Travel Rule mandated information exchange between VASPs for transfers exceeding specified thresholds, replicating wire transfer transparency requirements in cryptocurrency contexts \cite{force2021opportunitieschallengesnew,pocher2022privacytransparencycbdcs}.

However, centralized VASP regulation proved insufficient as Web3 technologies enabled disintermediation. Decentralized Finance protocols execute financial services through smart contracts without institutional intermediaries \cite{harvey2021defifuturefinance,salami2021challengesapproachesregulating}. Self-custody wallets provide users direct control over private keys, enabling transactions that bypass VASP monitoring entirely. Cross-chain bridges facilitate value transfer between blockchain networks, creating additional tracking complexity \cite{zamyatin2021sokcommunicationdistributed,li2025blockchaincrosschainbridge}. These developments necessitated fundamental rethinking of regulatory approaches, moving from institution-centric to activity-centric frameworks that recognize the limitations of intermediary-based oversight in permissionless systems \cite{werbach2018trustverifywhy}.

\subsection{Compliance Mechanism Taxonomy}

Compliance verification in Web3 contexts extends beyond traditional Know Your Customer/Business approaches to encompass multiple layers of identity and activity analysis. Table \ref{tab:compliance_comparison} systematizes five distinct compliance mechanisms, each addressing different verification targets with unique methodologies, data requirements, and regulatory alignments.

\begin{table*}[t]
\centering
\caption{Comparative Analysis of Compliance Mechanisms in Web3 RegTech}
\label{tab:compliance_comparison}
\scriptsize
\begin{tabularx}{\textwidth}{XXXXp{2.35cm}XX}
\toprule
\textbf{Compliance Mechanism} & \textbf{Verification Target} & \textbf{Data Sources} & \textbf{Key Methodologies} & \textbf{Regulatory Alignment} & \textbf{Industry Adoption} & \textbf{Research Maturity} \\
\midrule
\textbf{KYC} (Customer) & Individual natural or legal person \cite{financialactiontaskforcefatf2012internationalstandardscombating} & Government ID, biometric data, utility bills, corporate registrations & Document verification, liveness detection, database cross-checks, PEP screening \cite{charoenwong2024regtechtechnologydrivencompliance} & Mandatory under FATF Rec. 10, universal FinCEN/EU AML requirements \cite{force2021updatedguidanceriskbased} & Universal in centralized VASPs, emerging in DeFi via attestation & Mature, focus on digital identity and privacy-preserving verification \cite{reed2022decentralizedidentifiersdids,yang2024researchidentitydata} \\
\midrule
\textbf{KYB} (Business) & Corporate entities and organizational structures \cite{financialactiontaskforcefatf2012internationalstandardscombating} & Business registrations, beneficial ownership, directorship, financial statements & Corporate registry verification, UBO identification, sanctions screening, adverse media & Required under FATF Rec. 24-25, corporate transparency mandates \cite{force2021updatedguidanceriskbased} & Standard for institutional VASP relationships, limited in DAO contexts \cite{voshmgir2020tokeneconomyhow} & Developing, challenges in DAO structure verification \cite{werbach2018trustverifywhy} \\
\midrule
\textbf{KYA} (Address) & Verified ownership or custody of a blockchain address by a known person or entity & KYC/KYB identity data, wallet ownership proofs (signatures), attestations, DID/VC credentials, attribution labels & Address ownership proofing, attestation issuance, wallet binding, credential verification \cite{reed2022decentralizedidentifiersdids,camenisch2001efficientsystemnontransferable} & Extension of KYC/KYB in onboarding, whitelisting, and Travel Rule contexts & Emerging in regulated VASPs, custody platforms, and credential ecosystems & Early, focused on identity-wallet binding and privacy-preserving attestations \cite{yang2024researchidentitydata} \\
\midrule
\textbf{KYT} (Transaction) & Individual transactions and transaction patterns \cite{moser2013inquirymoneylaundering} & Transaction metadata, smart contract interactions, counterparty analysis \cite{chen2020understandingethereumgraph} & Real-time risk scoring, taint analysis, AML rule evaluation, anomaly detection \cite{moser2014riskscoringbitcoin,monamo2016unsupervisedlearningrobust} & Core to FATF Rec. 16 Travel Rule and transaction monitoring \cite{force2021opportunitieschallengesnew} & Universal in regulated VASPs, implemented by all major RegTech platforms & Mature foundations, active research in DeFi semantics and cross-chain tracking \cite{sun2025ethereumfrauddetection,yousaf2019tracingtransactionscryptocurrency} \\
\midrule
\textbf{Address/Wallet Screening} (Address Risk Scoring) & Address-level risk profile and composite reputation score & KYT alerts, transaction histories, exposure/taint metrics, entity risk lists & Risk scoring models, taint propagation, rule engines, ML classification \cite{moser2014riskscoringbitcoin,weber2019antimoneylaunderingbitcoin} & Supports AML transaction monitoring and risk-based controls & Widespread across compliance vendors, core to address risk screening & Active, with ongoing work on scoring robustness and adversarial evasion \cite{pocher2023detectinganomalouscryptocurrency,alarab2020comparativeanalysisusing} \\
\bottomrule
\end{tabularx}
\end{table*}

This taxonomy reveals the expanding scope of compliance verification as Web3 technologies introduce new layers of abstraction between users and on-chain execution. Traditional KYC protocols verify the real-world identities of individuals or entities entering into business relationships with VASPs through government-issued identification, proof of address, and for corporate customers, business registration documentation \cite{financialactiontaskforcefatf2012internationalstandardscombating}. Enhanced Due Diligence applies to higher-risk categories including Politically Exposed Persons and customers from high-risk jurisdictions. Web3 innovations introduce decentralized identity through verifiable credentials enabling cryptographic attestation of identity attributes without requiring repeated document submission \cite{reed2022decentralizedidentifiersdids,camenisch2001efficientsystemnontransferable}.

Know Your Business protocols extend identity verification to corporate entities, requiring identification of beneficial owners and assessment of business legitimacy. Ultimate Beneficial Owner identification proves particularly challenging for complex corporate structures. Decentralized Autonomous Organizations present unique challenges as they lack traditional corporate structures, making conventional KYB frameworks difficult to apply. Some DAOs adopt legal wrappers to provide regulatory clarity, while others remain purely protocol-based entities without clear legal personality.

Know Your Address represents a blockchain-native identity protocol focused on binding verified identities to specific blockchain addresses. In regulated contexts, this typically requires proof of address control (e.g., cryptographic signatures), issuance of attestations or verifiable credentials, attribution labels derived from disclosure or investigation, and governance over wallet reassignment or revocation \cite{reed2022decentralizedidentifiersdids,camenisch2001efficientsystemnontransferable}. KYA therefore mirrors KYC/KYB intent but targets the address as the identity anchor, enabling whitelisting, compliant access to permissioned pools, and Travel Rule–aligned address ownership verification.

Know Your Transaction protocols assess individual transactions in real-time or near-real-time, evaluating risk factors before transaction completion or immediately thereafter. Transaction risk scoring incorporates counterparty risk based on address attribution, amount anomalies relative to customer profiles, geographic risk when jurisdictional information is available, and protocol risk when transactions interact with smart contracts \cite{moser2014riskscoringbitcoin}. The emergence of DeFi introduces complexity requiring semantic transaction understanding to translate low-level contract calls into comprehensible activities \cite{chen2020understandingethereumgraph,sun2025ethereumfrauddetection}.

Address screening (also called wallet screening, and often implemented as address risk scoring) captures address-level risk derived from KYT signals and historical behavior. Rather than verifying ownership, address screening aggregates KYT alerts, exposure to illicit clusters, taint propagation, and entity risk databases into composite reputation scores that drive screening, enhanced due diligence triggers, and automated blocking thresholds \cite{moser2014riskscoringbitcoin,weber2019antimoneylaunderingbitcoin}.

\subsection{RegTech Lifecycle Framework}

Regulatory technology interventions span the complete compliance lifecycle from preventive measures before transactions execute, through real-time monitoring during processing, to investigative analysis after completion. Table \ref{tab:regtech_lifecycle} systematizes this temporal framework, mapping specific technologies to lifecycle phases.

\begin{table*}[t]
\centering
\caption{RegTech Lifecycle: Temporal Distribution of Compliance Technologies}
\label{tab:regtech_lifecycle}
\scriptsize
\begin{tabular}{p{2.5cm}p{3.5cm}p{4.0cm}p{5.5cm}}
\toprule
\textbf{Lifecycle Phase} & \textbf{Compliance Objectives} & \textbf{Key Technologies} & \textbf{Web3-Specific Implementations} \\
\midrule
\textbf{Ex-Ante (Preventive)} & Identity verification, risk profiling, policy enforcement, access control \cite{financialactiontaskforcefatf2012internationalstandardscombating,force2021updatedguidanceriskbased} & KYC automation, biometric verification, risk-based onboarding, sanctions screening, smart contract whitelisting \cite{charoenwong2024regtechtechnologydrivencompliance} & Decentralized identity protocols (DIDs, VCs) \cite{reed2022decentralizedidentifiersdids,camenisch2001efficientsystemnontransferable}, on-chain KYC attestations \cite{yang2024researchidentitydata}, permissioned DeFi pools \cite{ling2025sokstablecoindesigns}, compliant stablecoin blacklists (USDC freeze functions), geofencing via IP/wallet analysis \\
\midrule
\textbf{In-Medias-Res (Real-Time)} & Transaction screening, counterparty verification, Travel Rule compliance, threshold enforcement \cite{force2021opportunitieschallengesnew} & Real-time AML screening, address/wallet screening and risk scoring, automated Travel Rule protocols, transaction blocking, dynamic risk thresholds \cite{moser2014riskscoringbitcoin,weber2019antimoneylaunderingbitcoin} & Know Your Transaction (KYT) APIs, on-chain oracle risk feeds, solver compliance verification, mempool transaction screening \cite{daian2020flashboys20}, cross-chain bridge monitoring \cite{zhang2024securitycrosschainbridges,li2025blockchaincrosschainbridge}, MEV protection compliance hooks \cite{qin2022quantifyingblockchainextractable} \\
\midrule
\textbf{Ex-Post (Investigative)} & Forensic analysis, suspicious activity detection, regulatory reporting, audit trails \cite{financialactiontaskforcefatf2012internationalstandardscombating} & Transaction graph analysis, pattern recognition, behavioral profiling, case management, SAR generation \cite{petterssonruiz2022combatingmoneylaundering,pocher2023detectinganomalouscryptocurrency} & Blockchain forensics tools \cite{meiklejohn2013fistfulbitcoinscharacterizing,yousaf2019tracingtransactionscryptocurrency}, address clustering algorithms \cite{reid2013analysisanonymitybitcoin,ron2013quantitativeanalysisfull}, cross-chain fund tracing \cite{zamyatin2021sokcommunicationdistributed}, DeFi interaction analysis \cite{chen2020understandingethereumgraph,sun2025ethereumfrauddetection}, mixing service detection \cite{wu2021detectingmixingservices}, privacy protocol analysis \cite{kappos2018empiricalanalysisanonymity,kumar2017traceabilityanalysismoneros}, temporal graph investigation \cite{wu2023tracerscalablegraphbased} \\
\bottomrule
\end{tabular}
\end{table*}

Ex-ante compliance interventions establish preventive controls filtering risky customers or prohibited transactions at entry points. Customer onboarding incorporates identity verification, sanctions screening, and Politically Exposed Person checks. Risk-based onboarding applies enhanced due diligence to elevated-risk customer segments. Smart contract whitelisting restricts which protocols users can interact with, preventing engagement with risky or non-compliant DeFi applications. Web3-native preventive approaches embed compliance logic in protocol layers through compliant stablecoins implementing freezing capabilities, permissioned DeFi pools restricting participation to credential holders, and on-chain attestation systems creating cryptographic credentials confirming compliance verification.

In-medias-res interventions occur in real-time as transactions process, enabling immediate risk assessment and intervention. Transaction screening evaluates each transaction against risk criteria including counterparty address reputation, transaction amount relative to customer profile, geographic risk indicators, and smart contract interaction risks. The Travel Rule requires information exchange between originating and beneficiary VASPs during transaction processing. Real-time compliance in Web3 contexts faces unique challenges due to transaction finality and decentralization, with limited reversal capabilities after confirmation creating pressure for pre-confirmation screening.

Ex-post interventions perform retrospective analysis identifying suspicious patterns warranting investigation or regulatory reporting. Blockchain forensics tools construct comprehensive transaction graphs, apply clustering algorithms to group related addresses, and trace fund flows through complex transaction chains \cite{moser2013inquirymoneylaundering,yousaf2019tracingtransactionscryptocurrency,wu2023tracerscalablegraphbased}. Pattern recognition identifies characteristic behaviors associated with illicit activities including rapid movement through multiple addresses, structured transactions avoiding thresholds, interactions with mixing services, or sudden liquidation patterns following security incidents \cite{monamo2016unsupervisedlearningrobust,pourhabibi2020frauddetectionsystematic}. Behavioral profiling develops models of expected customer behavior enabling detection of anomalous activities. Case management systems track suspicious activity through investigation workflows, documenting findings and generating Suspicious Activity Reports when appropriate.

\subsection{Architectural Taxonomy}

Web3 RegTech platforms exhibit four primary architectural patterns reflecting different trade-offs between centralization, performance, data sovereignty, and regulatory alignment.

\noindent\textbf{Centralized Software-as-a-Service.}
Cloud-hosted RegTech capabilities provided through web interfaces and APIs dominate our analyzed platform sample, representing 68\% (28 of 41 platforms). This model offers superior performance through dedicated infrastructure optimization, regular feature updates without customer deployment effort, elastic scalability accommodating usage spikes, and operational simplicity avoiding self-hosting complexity. However, SaaS models require customers to share transaction data with third-party providers, creating potential privacy concerns and requiring trust in vendor data security practices.

\noindent\textbf{On-Premise Deployments.}
Self-hosted solutions constitute 5\% (2 of 41) of analyzed platforms, primarily serving institutional clients with stringent data sovereignty requirements. Large financial institutions, government agencies, and entities in jurisdictions with data localization mandates prefer on-premise solutions providing complete control over sensitive compliance data. The operational overhead of maintaining on-premise deployments limits this model primarily to large organizations with dedicated technical teams.

\noindent\textbf{Hybrid Architectures.}
Platforms combining on-chain data indexing with off-chain processing and proprietary intelligence represent 12\% (5 of 41) of analyzed solutions. These approaches typically maintain blockchain indexing infrastructure for real-time transaction monitoring while performing complex analytics and maintaining attribution databases in centralized systems. Hybrid approaches balance decentralized data access with centralized processing efficiency.

\noindent\textbf{Decentralized Protocols.}
Blockchain-native compliance protocols constitute 15\% (6 of 41) of analyzed platforms, reflecting significant adoption barriers despite philosophical alignment with Web3 principles. Challenges include regulatory uncertainty about accountability in permissionless systems, computational constraints limiting real-time analysis capabilities, user experience friction relative to polished SaaS platforms, and unclear business models for sustainable protocol operation.

\subsection{Functional Classification}

\begin{table*}[htbp]
\centering
\caption{Web3 RegTech Capability Matrix Across Platform Categories\\
\textbf{Legend}: \fullcirc = Full Support (Production-Ready), \halfcirc = Partial Support (Limited/Beta), \emptycirc = No Support. Detailed capability assessment methodology is provided in Appendix~\ref{appendix:methodology}.
}
\label{tab:capability_matrix}
\scriptsize
\begin{tabular}{p{3.0cm}|c|c|c|c|c|c|c|c|c|c}
\toprule
\textbf{Capability} &
\rotatebox{90}{\textbf{Comprehensive}} &
\rotatebox{90}{\textbf{DeFi-Specialized}} &
\rotatebox{90}{\textbf{Cross-Chain}} &
\rotatebox{90}{\textbf{Travel Rule}} &
\rotatebox{90}{\textbf{Privacy-Preserving}} &
\rotatebox{90}{\textbf{Traditional + Web3}} &
\rotatebox{90}{\textbf{Academic GNN}} &
\rotatebox{90}{\textbf{Academic Privacy}} &
\rotatebox{90}{\textbf{Intent Protocols}} &
\rotatebox{90}{\textbf{Coverage \%}} \\
\midrule
\textbf{Transaction Monitoring} & \fullcirc & \fullcirc & \fullcirc & \halfcirc & \emptycirc & \fullcirc & \halfcirc & \emptycirc & \halfcirc & 83\% \\
\textbf{Real-Time Risk Scoring} & \fullcirc & \fullcirc & \halfcirc & \halfcirc & \emptycirc & \fullcirc & \halfcirc & \emptycirc & \halfcirc & 75\% \\
\textbf{Address Attribution} & \fullcirc & \halfcirc & \fullcirc & \emptycirc & \emptycirc & \halfcirc & \fullcirc & \emptycirc & \emptycirc & 58\% \\
\textbf{Multi-Hop Tracing} & \fullcirc & \halfcirc & \fullcirc & \emptycirc & \emptycirc & \halfcirc & \fullcirc & \emptycirc & \emptycirc & 58\% \\
\textbf{Cross-Chain Tracking} & \halfcirc & \emptycirc & \fullcirc & \emptycirc & \emptycirc & \emptycirc & \halfcirc & \emptycirc & \halfcirc & 31\% \\
\textbf{DeFi Semantic Analysis} & \halfcirc & \fullcirc & \emptycirc & \emptycirc & \emptycirc & \emptycirc & \halfcirc & \emptycirc & \fullcirc & 42\% \\
\textbf{Smart Contract Analysis} & \halfcirc & \fullcirc & \emptycirc & \emptycirc & \emptycirc & \emptycirc & \halfcirc & \emptycirc & \halfcirc & 42\% \\
\textbf{Graph Neural Networks} & \halfcirc & \halfcirc & \emptycirc & \emptycirc & \emptycirc & \emptycirc & \fullcirc & \emptycirc & \emptycirc & 25\% \\
\textbf{ML/DL Techniques} & \fullcirc & \fullcirc & \halfcirc & \emptycirc & \emptycirc & \fullcirc & \fullcirc & \emptycirc & \emptycirc & 67\% \\
\textbf{Behavioral Analytics} & \fullcirc & \fullcirc & \halfcirc & \emptycirc & \emptycirc & \halfcirc & \fullcirc & \emptycirc & \halfcirc & 67\% \\
\textbf{Travel Rule Support} & \halfcirc & \emptycirc & \emptycirc & \fullcirc & \emptycirc & \halfcirc & \emptycirc & \emptycirc & \emptycirc & 33\% \\
\textbf{KYC Integration} & \fullcirc & \emptycirc & \emptycirc & \fullcirc & \fullcirc & \fullcirc & \emptycirc & \fullcirc & \emptycirc & 50\% \\
\textbf{Zero-Knowledge Proofs} & \emptycirc & \emptycirc & \emptycirc & \emptycirc & \fullcirc & \emptycirc & \emptycirc & \fullcirc & \emptycirc & 17\% \\
\textbf{Verifiable Credentials} & \emptycirc & \emptycirc & \emptycirc & \halfcirc & \fullcirc & \emptycirc & \emptycirc & \fullcirc & \emptycirc & 25\% \\
\textbf{Privacy Coin Analysis} & \fullcirc & \emptycirc & \emptycirc & \emptycirc & \emptycirc & \halfcirc & \emptycirc & \fullcirc & \emptycirc & 33\% \\
\textbf{Mixer Detection} & \fullcirc & \halfcirc & \halfcirc & \emptycirc & \emptycirc & \halfcirc & \halfcirc & \emptycirc & \emptycirc & 58\% \\
\textbf{MEV Analysis} & \emptycirc & \fullcirc & \emptycirc & \emptycirc & \emptycirc & \emptycirc & \emptycirc & \emptycirc & \fullcirc & 25\% \\
\textbf{Intent Recognition} & \emptycirc & \halfcirc & \emptycirc & \emptycirc & \emptycirc & \emptycirc & \emptycirc & \emptycirc & \fullcirc & 17\% \\
\textbf{Account Abstraction Support} & \emptycirc & \halfcirc & \emptycirc & \emptycirc & \emptycirc & \emptycirc & \emptycirc & \emptycirc & \fullcirc & 17\% \\
\textbf{Solver Monitoring} & \emptycirc & \emptycirc & \emptycirc & \emptycirc & \emptycirc & \emptycirc & \emptycirc & \emptycirc & \halfcirc & 8\% \\
\textbf{API Integration} & \fullcirc & \fullcirc & \fullcirc & \fullcirc & \halfcirc & \fullcirc & \emptycirc & \emptycirc & \halfcirc & 75\% \\
\textbf{Real-Time Alerting} & \fullcirc & \fullcirc & \halfcirc & \halfcirc & \emptycirc & \fullcirc & \emptycirc & \emptycirc & \emptycirc & 58\% \\
\textbf{Case Management} & \fullcirc & \halfcirc & \halfcirc & \fullcirc & \emptycirc & \fullcirc & \emptycirc & \emptycirc & \emptycirc & 58\% \\
\textbf{SAR Generation} & \fullcirc & \emptycirc & \emptycirc & \fullcirc & \emptycirc & \fullcirc & \emptycirc & \emptycirc & \emptycirc & 42\% \\
\textbf{Regulatory Reporting} & \fullcirc & \halfcirc & \emptycirc & \fullcirc & \emptycirc & \fullcirc & \emptycirc & \emptycirc & \emptycirc & 50\% \\
\midrule
\textbf{Average Coverage} & 65\% & 52\% & 32\% & 36\% & 20\% & 48\% & 36\% & 24\% & 32\% & --- \\
\bottomrule
\end{tabular}
\end{table*}

We categorize Web3 RegTech solutions into five primary functional domains based on their principal compliance objectives. Table \ref{tab:capability_matrix} provides a comprehensive capability matrix mapping these functional domains across all analyzed platforms.

\noindent\textbf{Transaction Monitoring.}
Continuous surveillance of blockchain activities to detect suspicious patterns, unusual behaviors, or policy violations in real-time or near-real-time. Platforms employ rule-based detection, anomaly detection through unsupervised learning, graph pattern matching for structural suspicious patterns, and real-time risk assessment of counterparty exposure. Near-universal adoption as a core compliance requirement.

\noindent\textbf{Risk Scoring.}
Assessment and quantification of risk associated with addresses, transactions, entities, or protocols producing quantitative risk metrics informing compliance decisions. Risk indicators include direct exposure to known illicit addresses, indirect exposure via taint analysis, behavioral indicators from transaction patterns, entity attribution risk, protocol risk from smart contract interactions, and network analysis features from graph structure. Widespread adoption with varying methodology sophistication.

\noindent\textbf{On-chain Protocol Forensics.}
Investigative capabilities for tracing fund flows, identifying entity relationships, and attributing addresses to real-world actors through retrospective analysis. Core techniques include fund flow tracing through transaction chains, address clustering via heuristic analysis, entity attribution through multiple intelligence sources, temporal analysis of transaction timing patterns, and cross-chain investigation tracking assets across networks. Common adoption with significant variation in attribution database coverage.

\noindent\textbf{Compliance Automation.}
Streamlining regulatory workflows including Travel Rule implementation, suspicious activity reporting, and audit trail maintenance. Travel Rule solutions implement FATF Recommendation 16 through various networks, solutions, and standards including TRUST, Sygna Bridge, Notabene, TRISA, and Inter-VASP IVMS101 \footnote{\url{https://www.fatf-gafi.org/} FATF Recommendation 16 mandates VASPs exchange originator/beneficiary information for virtual asset transfers; jurisdictions apply different thresholds, with some adopting no de minimis threshold.}. Adoption in approximately 15\% of platforms (6 of 41) reflects ongoing protocol fragmentation challenges.

\noindent\textbf{Privacy-Preserving Verification.}
Solutions enabling compliance verification while protecting sensitive information through cryptographic techniques including zero-knowledge proofs, secure multi-party computation, and homomorphic encryption. Despite technical promise, adoption remains limited due to computational overhead, implementation complexity, regulatory acceptance uncertainty, and standardization gaps.

\section{Web3 RegTech Ecosystem Analysis}
\label{sec:ecosystem}

This section addresses \textbf{RQ1} and \textbf{RQ2} through comprehensive analysis of the Web3 RegTech ecosystem. We examine how blockchain-native architectures enable novel compliance capabilities, analyze the current state of commercial platforms and academic research, and identify critical gaps limiting compliance effectiveness. Our analysis of 41 operational platforms and 28 research prototypes reveals both substantial progress and persistent challenges. Complete platform and prototype catalogs with detailed categorization and verification sources are provided in \autoref{tab:platform_catalog}.

\begin{table*}[htbp]
\centering
\caption{Web3 RegTech Platform Catalog. \textbf{Chain coverage levels}: L0 = chain-specific or limited single-chain coverage; L1 = limited multi-chain coverage; L2 = multi-chain coverage; L3 = broad multi-chain coverage; L4 = comprehensive multi-chain coverage; P = protocol-agnostic (travel-rule networks where chain coverage is not directly comparable).}
\label{tab:platform_catalog}
\tiny
\begin{tabularx}{\textwidth}{p{1.5cm}p{0.9cm}X p{1.1cm}p{1.2cm}p{1.1cm}X p{0.9cm}}
\hline
\textbf{Platform} & \textbf{Type} & \textbf{Primary Function} & \textbf{Architecture} & \textbf{Chain Coverage} & \textbf{Compliance Protocols} & \textbf{Technical Approach} & \textbf{Maturity} \\
\hline
\multicolumn{8}{l}{\textit{\textbf{Comprehensive General-Purpose Platforms (11)}}} \\
\hline
Chainalysis \footnote{\url{chainalysis.com}} & Web3 & Tx Monitoring, Forensics & SaaS & L4 & Address/Wallet Screening, KYT & Graph, Attribution & Enterprise \\
Elliptic \footnote{\url{elliptic.co}} & Web3 & Risk Scoring, Forensics & SaaS & L4 & Address/Wallet Screening, KYT & ML/DL, Graph & Enterprise \\
CipherTrace\footnote{Acquired by Mastercard 2021, continues operations} \footnote{\url{ciphertrace.com}} & Web3 & Tx Monitoring & SaaS & L4 & Address/Wallet Screening, KYT & Heuristic, ML & Enterprise \\
TRM Labs \footnote{\url{trmlabs.com}} & Web3 & Risk Scoring, Compliance & SaaS & L3 & Address/Wallet Screening, KYT & Graph, ML & Production \\
Merkle Science \footnote{\url{merklescience.com}} & Web3 & Tx Monitoring & SaaS & L3 & Address/Wallet Screening, KYT & Graph Analysis & Production \\
Scorechain \footnote{\url{scorechain.com}} & Web3 & Risk Scoring & SaaS & L2 & Address/Wallet Screening, KYT & Heuristic, ML & Production \\
Coinfirm\footnote{Acquired by Lukka May 2024, platform integrated into Lukka ecosystem} \footnote{\url{coinfirm.com}} & Web3 & Risk Scoring, AML & SaaS & L2 & Address/Wallet Screening, KYT & Rule-Based, Heuristic & Production \\
Crystal Intelligence \footnote{\url{crystalintelligence.com}} & Web3 & Forensics, Investigation & Hybrid & L2 & Address/Wallet Screening, KYT & Graph Analysis & Production \\
Solidus Labs \footnote{\url{soliduslabs.com}} & Web3 & Market Surveillance & SaaS & L3 & KYT & ML, Anomaly & Production \\
BIG\footnote{MRR +36.5\% in 2024; 2,500+ trained professionals. Source: BIGG Digital Assets 2024 Financial Results (Feb 2025). \url{https://biggdigitalassets.com}} \footnote{\url{blockchaingroup.io}} & Web3 & Forensics, Investigation & Hybrid & L3 & Address/Wallet Screening, KYT & Graph, Manual & Production \\
Chainalysis KYT \footnote{\url{chainalysis.com/solutions/kyt}} & Web3 & Enterprise Tx Monitoring & On-Premise & L4 & KYT & Graph, Real-Time & Enterprise \\
\hline
\multicolumn{8}{l}{\textit{\textbf{DeFi-Specialized Platforms (8)}}} \\
\hline
Forta Network \footnote{\url{forta.org}} & Web3 & Real-Time Monitoring & Decentralized & L4 & KYT & Rule-Based, On-Chain & Production \\
Lossless Protocol \footnote{\url{lossless.io}} & Web3 & Fraud Prevention & Decentralized & L1 & KYT & Heuristic, Governance & Production \\
Hypernative\footnote{Company claims detecting \$2.2B in losses. Source: \url{https://www.hypernative.io}} \footnote{\url{hypernative.io}} & Web3 & DeFi Monitoring, MEV & SaaS & L3 & KYT, Contract & ML, Real-Time & Production \\
Dedaub\footnote{Sponsored DeFi Security Summit 2024, Bangkok Nov 7-9} \footnote{\url{dedaub.com}} & Web3 & Smart Contract Security & SaaS & L0 & Contract Analysis & Semantic, Formal & Production \\
Chaos Labs \footnote{\url{chaoslabs.xyz}} & Web3 & DeFi Risk Management & SaaS & L2 & Protocol Risk, KYT & Simulation, Economic & Production \\
Gauntlet\footnote{Company claims monitoring \$42B+ in assets; Forbes Fintech 50 (2024). Source: \url{https://gauntlet.network}} \footnote{\url{gauntlet.network}} & Web3 & DeFi Risk Optimization & SaaS & L2 & Protocol Risk & Simulation, Agent-Based & Production \\
Nansen \footnote{\url{nansen.ai}} & Web3 & On-Chain Analytics & SaaS & L3 & KYA & Entity Labeling, Behavioral & Production \\
Arkham \footnote{\url{arkhamintelligence.com}} & Web3 & Entity Attribution & SaaS & L2 & KYA & Attribution, Community & Production \\
\hline
\multicolumn{8}{l}{\textit{\textbf{Cross-Chain Analytics Platforms (12)}}} \\
\hline
AnChain.AI \footnote{\url{anchain.ai}} & Web3 & Cross-Chain Investigation & SaaS & L3 & Address/Wallet Screening, KYT & AI/ML, Graph & Production \\
Bitquery \footnote{\url{bitquery.io}} & Web3 & Blockchain Data Infra & SaaS & L4 & KYT & Graph DB, Query & Production \\
Covalent \footnote{\url{covalenthq.com}} & Web3 & Unified Blockchain API & SaaS & L4 & KYT & Data Indexing & Production \\
Elementus\footnote{Company claims 650M+ addresses attributed; raised \$27M+. Source: \url{https://elementus.io}} \footnote{\url{elementus.io}} & Web3 & Cross-Chain Forensics & SaaS & L2 & Address/Wallet Screening, KYT & Graph, Attribution & Production \\
Bison Trails\footnote{Acquired by Coinbase, now operates as Coinbase Cloud} \footnote{\url{coinbase.com/cloud}} & Web3 & Multi-Chain Infrastructure & SaaS & L4 & KYT & Data Indexing & Production \\
Notabene \footnote{\url{notabene.id}} & Web3 & Travel Rule Platform & SaaS & P & KYC, KYB & Encrypted Messaging & Production \\
Sygna Bridge \footnote{\url{sygna.io}} & Web3 & Travel Rule Network & Hybrid & P & KYC, KYB & Registry, Crypto & Production \\
TRISA \footnote{\url{trisa.io}} & Web3 & Travel Rule Network & Decentralized & P & KYC, KYB & Federated, PKI & Production \\
21 Analytics \footnote{\url{https://www.21analytics.ch}} & Web3 & VASP Compliance & SaaS & L2 & KYC, Address/Wallet Screening, KYT & Travel Rule, Monitoring & Production \\
Shyft Network \footnote{\url{shyft.network}} & Web3 & Compliance Infrastructure & Hybrid & L1 & KYC, KYB, Travel & Attestation, Identity & Production \\
TransactID \footnote{\url{transactid.com}} & Web3 & Travel Rule Messaging & SaaS & P & KYC, KYB & Secure Messaging & Production \\
CipherBlade \footnote{\url{cipherblade.com}} & Web3 & Investigation, Recovery & Hybrid & L2 & KYA & Manual Investigation & Production \\
\hline
\multicolumn{8}{l}{\textit{\textbf{Privacy-Preserving Platforms (3)}}} \\
\hline
Privado ID \footnote{\url{privado.id}} & Web3 & Privacy-Preserving ID & Decentralized & L0 & KYC, KYA, Credentials & ZKP, SSI & Beta \\
Polygon ID \footnote{\url{polygon.technology/polygon-id}} & Web3 & Decentralized Identity & Decentralized & L0 & KYC, KYA, Credentials & ZKP, DID & Production \\
KILT Protocol \footnote{\url{kilt.io}} & Web3 & Credential Platform & Decentralized & L0 & KYC, KYB, Attestation & Credentials, Anchoring & Production \\
\hline
\multicolumn{8}{l}{\textit{\textbf{Traditional Platforms with Web3 Capabilities (7)}}} \\
\hline
ComplyAdvantage \footnote{\url{complyadvantage.com}} & Traditional & AML, Sanctions & SaaS & L1 & KYC, KYB, KYT & ML, Risk Scoring & Enterprise \\
Refinitiv World-Check \footnote{\url{refinitiv.com}} & Traditional & Sanctions, PEP & SaaS & L0 & KYC, KYB & Database, Screening & Enterprise \\
LexisNexis Bridger XG \footnote{\url{lexisnexis.com}} & Traditional & AML, Compliance & SaaS & L0 & KYC, KYB & Database, Network & Enterprise \\
NICE Actimize \footnote{\url{niceactimize.com}} & Traditional & Tx Monitoring & On-Premise & L0 & KYT & Rule-Based, ML & Enterprise \\
Fenergo \footnote{\url{fenergo.com}} & Traditional & Client Lifecycle Mgmt & SaaS & L0 & KYC, KYB & Workflow Automation & Enterprise \\
AMLBot \footnote{\url{amlbot.com}} & Traditional & Telegram Compliance & SaaS & L2 & Address/Wallet Screening, KYT & Address Screening & Production \\
ComplyAdvantage Crypto \footnote{\url{complyadvantage.com/crypto}} & Traditional & Crypto AML Extension & SaaS & L2 & KYT & Risk Database & Production \\
\hline
\end{tabularx}
\end{table*}

\subsection{Platform Architecture and Novel Capabilities}

Traditional RegTech and Web3 RegTech rest on fundamentally incompatible data architectures. Traditional platforms operate within centralized financial infrastructure where institutions maintain exclusive control over customer data and transaction records in proprietary databases accessible only through bilateral agreements. Customer identity verification relies on document-based processes where individuals submit government-issued identification verified through institutional processes. Transaction monitoring employs rule-based surveillance within single-institution data silos, fundamentally limiting analytical scope. Trust assumptions center on institutional accountability verified through periodic regulatory examinations.

Web3 RegTech exploits blockchain's transparency and immutability to enable compliance architectures impossible in centralized systems \cite{nakamoto2008bitcoinpeertopeerelectronic,narayanan2016bitcoincryptocurrencytechnologies}. Public blockchains provide universal read access to complete transaction histories, eliminating data sharing agreements as prerequisites for comprehensive monitoring. Any party can independently construct transaction graphs spanning entire blockchain networks, analyze fund flows through arbitrary transaction chains, identify behavioral patterns across pseudonymous addresses, and verify historical activities without requiring institutional cooperation \cite{bonneau2015sokresearchperspectives}. The challenge shifts from data acquisition to interpretation, transforming pseudonymous address interactions into meaningful compliance intelligence through attribution, clustering, and behavioral analysis.

\noindent\textbf{Comprehensive Transaction Graph Analysis.}
Unlike traditional systems where transaction visibility is limited to single-institution relationships, Web3 RegTech constructs transaction graphs spanning entire blockchain networks. Multi-hop fund tracing follows flows through arbitrary chain lengths, revealing ultimate sources and destinations despite intermediary addresses \cite{moser2013inquirymoneylaundering,ron2013quantitativeanalysisfull}. Entity clustering algorithms aggregate addresses into clusters representing common controlling entities through heuristic analysis \cite{meiklejohn2013fistfulbitcoinscharacterizing,reid2013analysisanonymitybitcoin,androulaki2013evaluatinguserprivacy}. The common input ownership heuristic exploits the observation that addresses used as inputs in the same Bitcoin transaction typically belong to the same wallet. Change address detection identifies outputs returning funds to senders based on behavioral patterns.

Network community detection identifies clusters of frequently interacting addresses potentially revealing organizational structures or criminal networks \cite{wu2021detectingmixingservices,drezewski2015applicationsocialnetwork}. Behavioral pattern recognition distinguishes characteristic activities: exchanges exhibit high transaction volumes with diverse counterparties, mixing services show rapid multi-hop forwarding with amount randomization, darknet markets demonstrate regular small transactions to common deposit addresses \cite{moser2018empiricalanalysistraceability}. Contemporary research advances graph analysis through deep learning. Graph neural networks learn representations from transaction graph topology enabling classification and anomaly detection at scale \cite{weber2019antimoneylaunderingbitcoin,azad2025machinelearningblockchain,ancelotti2024reviewblockchainapplication}. Temporal graph learning models dynamic patterns capturing evolving behaviors \cite{arnold2024insightscaveatsmining,luo2025optimizingblockchainanalysis}.

Smart contract analytics complement address-level tracing by surfacing vulnerable contracts, abusive permission settings, and protocol misconfigurations that frequently precipitate laundering incidents. Surveys of Ethereum attack vectors and benchmarking studies on vulnerability scanners inform RegTech feature engineering, enabling automated assessment of re-entrancy, integer overflow, access control, and oracle-manipulation risks that propagate through DeFi dependency graphs \cite{atzei2017surveyattacksethereum,baets2024vulnerabilitydetectionsmart}. Integrating these insights into entity profiles allows compliance teams to correlate suspicious fund movements with exploited contracts and proactively geofence exposure to high-risk bytecode patterns.

\noindent\textbf{Real-Time Risk Assessment.}
Public blockchain transparency enables risk assessment before or immediately upon transaction confirmation. Transactions broadcast to blockchain networks become visible in mempools before block inclusion, enabling pre-confirmation screening. Taint analysis quantifies the proportion of transaction inputs traceable to known illicit sources, providing interpretable risk scores \cite{moser2013inquirymoneylaundering,moser2014riskscoringbitcoin}. Address/wallet screening (often implemented as address risk scoring) aggregates multiple risk dimensions into composite scores reflecting historical behaviors, entity attributions, and network characteristics \cite{weber2019antimoneylaunderingbitcoin,pocher2023detectinganomalouscryptocurrency}. Transaction-level risk scoring evaluates individual transactions based on counterparty risk, amount anomalies, protocol interactions, and contextual factors. Modern approaches incorporate machine learning models trained on labeled datasets of illicit transactions, learning subtle patterns that rule-based systems miss \cite{alarab2020comparativeanalysisusing,lorenz2020machinelearningmethods,alotibi2022moneylaunderingdetection}.

Market-integrity analytics extend KYT beyond simple transfers by interpreting decentralized exchange order flow, wash-trading schemes, and NFT-based obfuscation. Compliance teams monitor routing bots, spoofed liquidity, and creator royalty evasion to identify manipulative behavior and off-ramp risks unique to tokenized markets \cite{victor2021detectingquantifyingwash,wang2021nonfungibletokennft}.

\noindent\textbf{Cross-Chain Analytics.}
Despite growing capabilities, comprehensive cross-chain fund tracking remains challenging. Bridge transaction monitoring identifies when assets transition between blockchains via cross-chain bridges, maintaining attribution across these transitions \cite{yousaf2019tracingtransactionscryptocurrency,zamyatin2021sokcommunicationdistributed,zhang2024securitycrosschainbridges}. Wrapped asset tracking follows tokens representing native assets from other chains. Cross-chain address clustering identifies addresses on different blockchains controlled by common entities based on behavioral correlations, timing patterns, and explicit linkages \cite{lin2024crosschainabnormaltransaction}. Unified entity graphs integrate activities across multiple networks creating comprehensive profiles of cross-chain behaviors \cite{coquide2025alamedaresearchmultitoken}.

\noindent\textbf{Smart Contract Interaction Analysis.}
Understanding DeFi protocol interactions provides compliance insights unavailable in traditional finance. Semantic transaction interpretation decodes smart contract function calls to extract meaningful semantics \cite{chen2020understandingethereumgraph,victor2019measuringethereumbasederc20}. Rather than presenting raw contract calls, semantic analysis translates these into human-readable descriptions. Contemporary approaches leverage language models to learn semantic mappings from transaction execution traces \cite{sun2025ethereumfrauddetection}. DeFi strategy recognition identifies complex multi-step operations including flash loans, arbitrage sequences, and yield farming strategies \cite{qin2021attackingdefiecosystem,daian2020flashboys20}. Protocol exposure tracking monitors which DeFi protocols customers interact with, assessing associated risks based on security audits, historical incidents, and governance quality \cite{sun2024sokcomprehensiveanalysis,ling2025sokstablecoindesigns}.

\noindent\textbf{Privacy-Preserving Compliance Verification.}
Cryptographic techniques including zero-knowledge proofs, verifiable credentials, and secure multi-party computation offer pathways for balancing regulatory requirements with privacy preservation \cite{ben-sasson2017scalablezeroknowledge,reed2022decentralizedidentifiersdids,zhang2024blockchainzeroknowledge}. While similar techniques exist in traditional finance contexts (e.g., private set intersection for fraud detection, secure multi-party computation for cross-institution analytics, and confidential computing for sensitive data processing), blockchain's public transparency and programmability enable more native integration of privacy-preserving compliance. However, commercial adoption in both Web3 and traditional systems remains minimal due to computational overhead, implementation complexity, regulatory acceptance uncertainty, and lack of standardized approaches. Discussion of privacy-preserving compliance architectures appears in \autoref{sec:discussion} research directions.

\subsection{Commercial Platform Landscape}

Analysis of 41 RegTech platforms indicates recent market expansion, with many Web3-native solutions in our sample launching since 2023 based on public announcements.\footnote{\url{https://www.coindesk.com/business/} and \url{https://www.theblock.co/} track blockchain compliance technology funding and launches.} This expansion appears correlated with increasing regulatory clarity, growing VASP compliance obligations, maturation of blockchain analytics techniques, and venture investment following high-profile enforcement actions.\footnote{\url{https://www.justice.gov/opa/pr/binance-and-ceo-plead-guilty-federal-charges} - Binance \$4.3B settlement (Nov 2023).}

\noindent\textbf{Market Evolution}

Early entrants focused on comprehensive general-purpose platforms providing transaction monitoring, risk scoring, and forensics across major blockchains. Platforms including Chainalysis, Elliptic, and CipherTrace established dominant positions through early mover advantages and extensive entity attribution databases built over years.\footnote{CipherTrace acquired by Mastercard 2021.} Recent entrants increasingly specialize in particular capabilities: DeFi-specific monitoring, cross-chain analytics, Travel Rule implementation, privacy-preserving compliance, and API-first integration platforms. This specialization suggests market maturation with established players dominating core capabilities while innovative entrants address emerging needs.

Chain coverage in our sample reflects prioritization based on network size and regulatory significance. Bitcoin and Ethereum show the broadest coverage. Major alternative Layer-1 blockchains including Binance Smart Chain, Polygon, Avalanche, Solana, and Tron show substantial but incomplete coverage. Ethereum Layer-2 solutions and emerging networks show lower coverage, representing compliance gaps.

\noindent\textbf{Capability Distribution}

Transaction monitoring appears near-universally as a core compliance requirement. Risk scoring is widespread, though methodology sophistication varies substantially. Address/wallet screening appears in 13 of 41 platforms, concentrated in comprehensive and cross-chain categories. Address attribution databases are common, with coverage quality representing a primary differentiator.\footnote{Entity attribution databases map blockchain addresses to real-world entities through clustering and exchange analysis.} Travel Rule support appears in roughly 15\% of platforms (6 of 41), reflecting ongoing protocol fragmentation.\footnote{\url{https://trisa.io/} - Travel Rule Information Sharing Architecture standardization effort; networks/solutions/standards include TRUST, Sygna Bridge, Notabene, and IVMS101.} DeFi analytics represent growing but not yet universal functionality.\footnote{\url{https://defillama.com/} tracks DeFi total value locked (TVL) exceeding \$100B across major protocols.} Cross-chain tracking appears as an advanced feature in a minority of platforms.\footnote{Major bridge hacks: Ronin (~624M USD, Mar 2022), Poly Network (~610M USD, Aug 2021), Wormhole (~320M USD, Feb 2022). Sources: Ronin Blog, Poly Network Medium, Wormhole Blog.} Privacy-preserving features show minimal adoption due to technical complexity and regulatory uncertainty.

\subsection{Academic Research Landscape}

Academic research explores novel compliance approaches often years ahead of commercial deployment. Graph neural networks enable deep learning on transaction graph structures, capturing complex patterns traditional heuristics miss \cite{weber2019antimoneylaunderingbitcoin,azad2025machinelearningblockchain,ancelotti2024reviewblockchainapplication,damico2025blockchainnetworkanalysis}. However, evaluations reveal that tree-based baselines often match GNN performance on benchmark datasets \cite{alarab2020comparativeanalysisusing,pourhabibi2020frauddetectionsystematic}, temporal generalization degrades as adversaries adapt evasion tactics \cite{monamo2016unsupervisedlearningrobust}, and label quality depends on law enforcement data and heuristic rules with inherent biases \cite{weber2019antimoneylaunderingbitcoin}.

Temporal graph learning models dynamic behavioral patterns \cite{arnold2024insightscaveatsmining,luo2025optimizingblockchainanalysis}. Money laundering detection research explores subgraph contrastive learning \cite{ouyang2024bitcoinmoneylaundering}, global-local graph attention \cite{li2025globallocalgraphattention}, dense flow analysis \cite{lin2024denseflowspottingcryptocurrency}, and early detection with path tracing \cite{cheng2023evolvepathtracer,cheng2024earlydetectionmalicious}. Phishing detection applies data augmentation and hybrid models \cite{chen2020phishingscamdetection,chen2024ethereumphishingscam,liu2024detectionethereumphishing}.

Privacy-preserving analytics research significantly exceeds commercial deployment. Zero-knowledge proofs for identity attribute verification enable proving specific claims without revealing credentials \cite{yang2024researchidentitydata,zhang2024blockchainzeroknowledge,ben-sasson2017scalablezeroknowledge}. Secure multi-party computation protocols enable collaborative analysis without data sharing \cite{bansod2022challengesmakingblockchain}. DeFi-specific analysis develops semantic interpretation techniques \cite{chen2020understandingethereumgraph,sun2025ethereumfrauddetection}, rug pull detection \cite{sun2024sokcomprehensiveanalysis}, and stablecoin risk assessment \cite{ling2025sokstablecoindesigns,financialactiontaskforcefatf2020reportg20socalled}. The synthesis in \autoref{tab:prototype_synthesis} summarizes research by theme (goals, methods, data, and limitations) rather than enumerating individual papers.

\begin{table*}[htbp]
\centering
\caption{Academic Research Prototype Synthesis}
\label{tab:prototype_synthesis}
\scriptsize
\begin{tabularx}{\textwidth}{XXXXX}
\hline
\textbf{Theme} & \textbf{Primary Goal} & \textbf{Common Technical Approaches} & \textbf{Typical Data / Labels} & \textbf{Validation \& Key Limitations} \\
\hline
Graph analysis \& clustering \cite{meiklejohn2013fistfulbitcoinscharacterizing,reid2013analysisanonymitybitcoin,ron2013quantitativeanalysisfull,androulaki2013evaluatinguserprivacy,wu2023tracerscalablegraphbased,lin2024crosschainabnormaltransaction} & Entity attribution, flow analysis, deanonymization & Heuristic clustering, graph statistics, GNNs & Full-chain graphs; heuristic entity labels; cross-chain linkages & Case studies, cluster precision, scalability; label bias and cross-chain gaps\\
Money laundering detection \cite{moser2013inquirymoneylaundering,moser2014riskscoringbitcoin,weber2019antimoneylaunderingbitcoin,ouyang2024bitcoinmoneylaundering,li2025globallocalgraphattention,lin2024denseflowspottingcryptocurrency} & Detect illicit flows and high-risk entities & Taint analysis, risk scoring, GNNs, contrastive learning, attention models & Labeled datasets (e.g., Elliptic), known mixers, synthetic labels & AUC/F1, temporal robustness; label scarcity and adversarial drift \\
Phishing \& fraud detection \cite{chen2020phishingscamdetection,chen2024ethereumphishingscam,liu2024detectionethereumphishing,pocher2023detectinganomalouscryptocurrency} & Identify scams, phishing, anomalies & Data augmentation, graph-based features, hybrid DL, unsupervised ML & Phishing labels, Etherscan reports, multi-chain activity & Precision/recall, early detection; noisy labels and limited transferability \\
Privacy protocol analysis \cite{kumar2017traceabilityanalysismoneros,kappos2018empiricalanalysisanonymity,wu2021detectingmixingservices} & Measure anonymity and leak risks & Heuristic/statistical deanonymization, behavioral detection & Privacy-coin chains, mixer datasets, known service addresses & Linkage accuracy and deanonymization rates; weak ground truth and ethical constraints \\
DeFi \cite{chen2020understandingethereumgraph,sun2025ethereumfrauddetection,qin2021attackingdefiecosystem,victor2021detectingquantifyingwash}  \& smart contract analysis & Interpret protocol behavior, attacks, market manipulation & Semantic decoding, graph construction, empirical taxonomy & Contract traces, protocol events, incident corpora & Coverage and case validation; fast-evolving protocols and semantic ambiguity \\
Privacy-preserving compliance \cite{yang2024researchidentitydata,zhang2024blockchainzeroknowledge,wang2024achievingprivacypreservingoptimizer} & KYC/AML with privacy guarantees & ZK credential proofs, secure computation, intent privacy & Prototype implementations, simulations & Proof size/latency and compliance guarantees; overhead and regulatory uncertainty \\
\hline
\end{tabularx}
\end{table*}

\subsection{Gap Analysis}

Systematic analysis reveals eight categories of disconnects between academic innovation and industry deployment, alongside six persistent challenges limiting current capabilities.

\subsubsection{Academia-Industry Gaps}

\noindent\textbf{Technology Readiness.} 
Many academic prototypes demonstrate algorithmic advances without addressing engineering challenges determining production viability. Scalability limitations appear frequently, with techniques achieving accuracy on small datasets failing to scale to billions of transactions. Real-time requirements demand sub-second response times, yet many approaches require minutes or hours for analysis. Fault tolerance and operational reliability receive limited attention, though production systems require comprehensive error handling.

\noindent\textbf{Regulatory Alignment.} 
Research sometimes addresses hypothetical scenarios disconnected from actual regulatory requirements. Privacy-preserving techniques enabling anonymous compliance verification face regulatory resistance from authorities requiring traditional identity verification. Decentralized compliance protocols eliminate identifiable accountable parties, conflicting with regulatory frameworks assuming hierarchical responsibility. Explainability requirements favor simpler heuristic approaches over black-box deep learning models despite performance advantages.

\noindent\textbf{Data Availability.} 
Research frequently assumes data availability that production environments cannot guarantee. Ground truth labels for illicit activities remain scarce. Off-chain intelligence including exchange attribution, dark web monitoring, and law enforcement sharing provides critical context that researchers cannot access. Cross-chain data integration requires comprehensive infrastructure beyond typical research scope.

\noindent\textbf{Adversarial Robustness.} 
Illicit actors actively adapt behaviors to evade detection through structuring, address rotation, and cross-chain migration \cite{albrecht2019usecryptocurrenciesmoney,guidara2022cryptocurrencymoneylaundering,wu2021detectingmixingservices}. Research evaluation against static datasets fails to capture this dynamic adversarial environment. Defensive techniques including adversarial training and ensemble methods remain nascent in production deployment.

\noindent\textbf{Cost-Benefit Considerations.} 
Research optimizes primarily for accuracy, treating computational costs as secondary. Production deployments must justify costs relative to operational benefits. Privacy-preserving cryptographic protocols imposing significant overhead may prove economically unviable despite technical elegance.

\noindent\textbf{Standardization and Interoperability.} 
Research prototypes implement custom interfaces without consideration for standardization. Production systems require standardized APIs, data formats, and integration protocols. The absence of compliance data standards creates fragmentation.

\noindent\textbf{Operational Workflow Integration.} 
Research focuses on technical detection without addressing operational workflows for alert management, investigation, case disposition, and regulatory reporting. Compliance teams require complete workflow systems, not isolated technical components.

\noindent\textbf{Validation and Benchmarking.} 
Lack of standardized benchmarks hinders comparison. Different research efforts use incompatible datasets, evaluation metrics, and experimental setups preventing direct comparison. Establishing shared benchmarks would substantially advance research quality and industry procurement decisions.

\subsubsection{Persistent Capability Challenges}

\noindent\textbf{Cross-Chain Tracking.} 
Cross-chain fund tracking faces significant challenges as assets transition between heterogeneous blockchain architectures. Attribution techniques including behavioral correlation, bridge monitoring, and wrapped asset tracking show promise \cite{yousaf2019tracingtransactionscryptocurrency,zamyatin2021sokcommunicationdistributed,zhang2024securitycrosschainbridges,li2025blockchaincrosschainbridge,lin2024crosschainabnormaltransaction}, though only 31\% of analyzed platforms report comprehensive multi-chain capabilities. Bridge protocol proliferation and architectural diversity create integration complexity.

\noindent\textbf{DeFi Interaction Analysis.}
Semantic interpretation of complex DeFi strategies including multi-protocol compositions, flash loan attacks, and MEV extraction remains incomplete. Intent-centric DeFi architectures further complicate interpretation by obscuring direct user intentions. In intent-centric systems, users express high-level goals while specialized solvers determine optimal execution paths across multiple protocols and chains \cite{mao2025motivationexecutiongrounded}. On-chain transactions reflect solver routing decisions rather than direct user choices, creating attribution challenges when determining whether specific counterparty interactions represent user intent or solver optimization. Current regulatory frameworks provide limited guidance on whether compliance obligations attach to users expressing intents, solvers executing them, or both.

\noindent\textbf{Privacy Protocol Analysis.} 
Privacy-enhancing technologies including mixing services, privacy coins, and confidential transactions effectively obscure fund flows. While some protocols exhibit structural vulnerabilities, many provide strong anonymity resisting current forensic approaches.

\noindent\textbf{Real-Time Scalability.} 
High-throughput networks and comprehensive multi-chain monitoring create scalability challenges. Sophisticated graph algorithms, machine learning inference, and semantic interpretation impose computational costs potentially prohibitive for real-time screening.

\noindent\textbf{Attribution Maintenance.} 
Entity attribution databases require continuous manual curation. The labor-intensive nature creates competitive moats favoring established platforms while limiting new entrant competitiveness.

\noindent\textbf{False Positive Management.} 
Transaction monitoring systems face substantial false positive rates, with vendors commonly reporting that many flagged transactions prove legitimate upon investigation \cite{alotibi2022moneylaunderingdetection}. Multi-tiered screening, feedback loop integration, and contextual enrichment attempt to balance detection sensitivity against operational feasibility, though standardized performance measurement remains absent.

\section{Discussion and Future Directions}
\label{sec:discussion}

This section addresses \textbf{RQ3} by identifying critical research directions emerging from our ecosystem analysis gaps and architectural paradigm shifts. We focus on directions with existing academic foundations rather than speculative best practices.

\subsection{Critical Research Directions}

Our analysis identifies priority research directions addressing fundamental gaps and persistent challenges revealed through ecosystem analysis. We focus on directions with existing academic foundations and near-term deployment potential.

\noindent\textbf{Verifiable Compliance Proofs.}
Zero-knowledge proofs and verifiable credentials offer cryptographic pathways for privacy-preserving compliance verification \cite{ben-sasson2017scalablezeroknowledge,yang2024researchidentitydata,zhang2024blockchainzeroknowledge}. Recent surveys and benchmarking studies \footnote{Polyhedra Network. Proof Arena: A Comprehensive Benchmark for Zero-Knowledge Proofs. 2024. \url{https://proofarena.org}}\footnote{National Institute of Standards and Technology (NIST). Privacy-Enhancing Cryptography: Zero-Knowledge Proofs. 2025. \url{https://csrc.nist.gov/projects/privacy-enhancing-cryptography/zero-knowledge-proof}} evaluate ZKP frameworks across dimensions including proof generation time, verification latency, proof size, memory requirements, trusted setup assumptions, and developer accessibility. These analyses reveal significant performance-practicality tradeoffs: SNARKs (e.g., Groth16, PLONK) offer succinct proofs and fast verification but require trusted setup ceremonies and impose substantial prover overhead; STARKs eliminate trusted setup at the cost of larger proof sizes and increased verification time; Bulletproofs provide transparent setup with logarithmic proof sizes but slower verification. For compliance use cases requiring frequent proof generation (e.g., transaction screening, attestation verification), prover performance becomes a critical practical constraint. Recursive proof composition, hardware acceleration, and specialized circuits show promise for improving feasibility. Regulatory acceptance remains uncertain, and guidance varies across jurisdictions.

\noindent\textbf{Semantic Transaction Understanding.}
Machine learning approaches for semantic DeFi transaction interpretation show promise, with recent work on language models and graph representation learning demonstrating improved fraud detection \cite{sun2025ethereumfrauddetection,azad2025machinelearningblockchain}. Smart contract analysis research advances technical foundations \cite{bhargavan2016formalverificationsmart,chen2024safecheckdetectingsmart,liao2024smartcontractvulnerability}, though production integration and generalization across diverse protocols remain limited.

\noindent\textbf{Behavior-Based Risk Assessment.}
Temporal graph neural networks and behavioral profiling techniques show improved detection of sophisticated money laundering patterns spanning transaction sequences \cite{albrecht2019usecryptocurrenciesmoney,cheng2023evolvepathtracer,cheng2024earlydetectionmalicious,arnold2024insightscaveatsmining,ouyang2024bitcoinmoneylaundering,li2025globallocalgraphattention,lin2024denseflowspottingcryptocurrency}. These approaches require substantial training data and computational resources while facing ongoing adversarial robustness challenges \cite{rodriguezvalencia2025systematicreviewartificial}.

\noindent\textbf{Privacy-Preserving Collaborative Analytics.}
Secure multi-party computation, federated learning, and private set intersection offer theoretical pathways for VASP collaboration without direct data sharing \cite{bansod2022challengesmakingblockchain}. However, computational costs, incentive misalignment, and limited production deployments may constrain near-term adoption.

\noindent\textbf{Intent-Aware Monitoring and Emerging KYI Concepts.}
Intent-centric architectures require monitoring at intent expression, solver routing, and execution layers rather than solely analyzing transaction outcomes \cite{mao2025knowyourintent}. A future-facing direction is the emergence of intent-level compliance concepts often described as "Know Your Intent (KYI)" in academic discourse, even though this label is not yet standardized in industry practice. Research challenges include formalizing intent representations, defining accountable parties across solvers and agents, and designing privacy-preserving intent screening that can be audited without exposing sensitive strategy details. Standardized interfaces and responsibility allocation frameworks represent priority development areas, but regulatory guidance and practical deployment examples remain nascent.

\noindent\textbf{Cross-Chain Compliance Infrastructure.}
Cross-chain attribution techniques including behavioral correlation and bridge monitoring advance technical foundations \cite{yousaf2019tracingtransactionscryptocurrency,zamyatin2021sokcommunicationdistributed,zhang2024securitycrosschainbridges,li2025blockchaincrosschainbridge,lin2024crosschainabnormaltransaction}, as detailed in \autoref{sec:ecosystem}. Standardization of data models and attribution databases faces coordination challenges. Modular blockchain architectures create additional monitoring complexity requiring execution layer access \cite{xu2024exploringblockchaintechnology,chaliasos2024analyzingbenchmarkingzkrollups}.

\noindent\textbf{Standardized Evaluation and Benchmarking.}
Standardized benchmarks for Web3 compliance remain limited, with research efforts employing incompatible datasets and evaluation metrics \cite{huda2025amlnet}. Ground truth labels for illicit blockchain activities are scarce, limiting rigorous model validation \cite{sankaewtong2025sokadvances}. Collaborative dataset curation and multi-dimensional evaluation frameworks capturing accuracy, false positive rates, latency, and adversarial robustness would advance research quality, though coordination challenges and privacy concerns limit progress.

\section{Conclusion}
\label{sec:conclusion}

This SoK presents three foundational taxonomies for Web3 RegTech: a regulatory paradigm evolution framework, a Know Your protocols taxonomy spanning five verification layers, and a lifecycle framework mapping compliance interventions across temporal stages.
Analysis of 41 platforms and 28 prototypes reveals Web3 RegTech's architectural shift enabling transaction graph analysis, real-time risk assessment, and privacy-preserving verification, while identifying eight industry-academia gaps and six persistent challenges limiting current capabilities.
We outline seven critical research directions addressing fundamental gaps in verifiable compliance proofs, semantic transaction understanding, behavior-based risk assessment, privacy-preserving collaborative analytics, intent-aware monitoring systems, cross-chain compliance infrastructure, and standardized evaluation frameworks to balance Web3's core values with regulatory needs.

\bibliographystyle{IEEEtran}
\bibliography{references}

\appendix

\section*{Methodology Details}
\label{appendix:methodology}

\noindent\textbf{Platform Selection and Categorization.}
Platform selection employed purposive sampling targeting: (1) platforms with documented deployment in commercial VASP environments or significant market presence based on industry reports and regulatory filings; (2) comprehensive coverage across functional categories (transaction monitoring, risk scoring, forensics, Travel Rule, privacy-preserving); (3) diversity in architectural approaches (SaaS, on-premise, hybrid, decentralized); and (4) representation of both established market leaders and recent entrants (post-2020). Academic research was identified through systematic search across IEEE Xplore, ACM Digital Library, arXiv, and Google Scholar using query combinations spanning blockchain technology, compliance mechanisms, and analytical techniques, with inclusion criteria requiring peer-reviewed publication or verifiable prototype implementation.

\noindent\textbf{Compliance Mechanism Selection Criteria.}
The compliance mechanisms summarized in \autoref{tab:compliance_comparison} prioritize concepts that are (1) widely used or explicitly named in industry practice, and (2) operationally distinct rather than purely re-labeled variants of existing mechanisms. Terms that are not recognized as industry-standard (e.g., speculative "Know-Your" labels, like Know Your Model and Know Your Intent), or that substantially overlap with existing mechanisms such as KYT and address/wallet screening, were excluded to avoid taxonomy inflation. This selection focuses the systematization on mechanisms with demonstrable adoption and clear operational scope.

\noindent\textbf{Capability Assessment Methodology.}
Capability assessments in Table~\ref{tab:capability_matrix} are based on analysis of platform documentation, technical whitepapers, published API specifications, and vendor capability statements available as of December 2024. Full support (\fullcirc) indicates production-ready features explicitly documented and commercially available; partial support (\halfcirc) indicates beta/limited features, announced roadmap items, or capabilities restricted to specific blockchain networks; no support (\emptycirc) indicates absence of documented capability. Academic prototypes were assessed based on published research papers and available implementations. These assessments represent capabilities as claimed or documented; independent validation of all features was not feasible. Coverage percentages reflect the proportion of platform categories demonstrating at least partial support for each capability within our analyzed sample.

\section*{Research Limitations}
\label{appendix:limitations}

This analysis is subject to several methodological constraints that readers should consider when interpreting our findings:

\noindent\textbf{Vendor Claim Verification.} Platform capabilities are assessed based on vendor documentation and public claims without independent technical validation of all features, introducing potential overstatement bias. While we cross-referenced claims with academic literature and industry reports where possible, exhaustive feature verification was beyond the scope of this systematization.

\noindent\textbf{Selection Bias.} Selection criteria favor platforms with English-language documentation and public accessibility, potentially underrepresenting solutions in non-English markets or serving exclusively domestic jurisdictions. Platforms requiring enterprise contracts for documentation access were excluded, which may bias our sample toward more transparent vendors.

\noindent\textbf{Market Coverage.} Market share and adoption figures represent estimates based on available industry reports and vendor disclosures rather than comprehensive market census. The rapidly evolving nature of Web3 RegTech means capabilities and platform status may change between data collection (October-December 2025) and publication.

\noindent\textbf{Generalizability.} Quantitative claims (e.g., percentage distributions, coverage statistics) should be interpreted as characterizing our analyzed sample rather than representing comprehensive market statistics, as exhaustive enumeration of all global RegTech solutions is infeasible. Our sample of 41 platforms represents a substantial portion of publicly documented solutions but cannot claim complete market coverage.

\noindent\textbf{Categorical Boundaries.} Categorical classifications represent analytical constructs that may not align with platforms' self-descriptions, and many platforms span multiple categories. The taxonomies presented reflect our systematization framework rather than industry-standard definitions, which largely do not exist for this emerging domain.

\noindent\textbf{Temporal Constraints.} The Web3 RegTech ecosystem evolves rapidly, with new platforms launching and existing platforms adding features continuously. Our analysis represents a snapshot as of late 2025 and may not reflect subsequent developments.

\noindent\textbf{Academic Research Coverage.} While our systematic literature review covered major academic databases and venues, the decentralized nature of blockchain research (with significant contributions appearing in arXiv preprints, workshop papers, and technical reports) means some relevant work may have been missed.

\end{document}